# A Hybrid Josephson Transmission Line and Passive Transmission Line Routing Framework for Single Flux Quantum Logic

Shucheng Yang, Xiaoping Gao, Ruoting Yang, Jie Ren, and Zhen Wang

*Abstract*—The Single Flux Quantum (SFQ) logic family is a novel digital logic as it provides ultra-fast and energy-efficient circuits. For large-scale SFQ circuit design, specialized electronic design automation (EDA) tools are required due to the differences in logic type, timing constraints and circuit architecture, in contrast to the CMOS logic. In order to improve the overall performance of an SFQ circuit, an efficient routing algorithm should be applied during the layout design to perform accurate timing adjustment for fixing hold violations and optimizing critical paths. Thus, a hybrid Josephson transmission line and passive transmission line routing framework is proposed. It consists of four main modules and an exploration of the potential timing performance based on the given layout placement. The proposed routing tool is demonstrated on seven testbench circuits. The obtained results demonstrate that the operating frequency is greatly improved, and all the hold violations are eliminated for each circuit.

## I. INTRODUCTION

Single Flux Quantum (SFQ) logic [1] is a digital logic based on Josephson junction and superconducting material. It is recently widely used as the superconducting fabrication technologies evolve. The Rapid-SFQ (RSFQ), its derived version - Energy-efficient RSFQ (ERSFQ)[2-5] and eSFQ [6] have large advantages on the operating frequency and power consumption, compared with the CMOS logic, due to the fact that the essential active element Josephson junction transmits SFQ signal on picosecond level and dissipates almost $10^{-19}$ J per switching [1]. In the RSFQ logic, the logic value "true" or "false" is defined by whether a flux quantum is stored in the superconducting loop. Therefore, the basic boolean gates - AND, OR and NOT, require an additional clock port for timing. This leads to a different timing strategy and clock tree structure for logic synthesis and layout synthesis.

For large-scale SFQ circuit design, the electronic design automation (EDA) tools are essential, due to the fact that a designer cannot manually perform logic synthesis and layout synthesis on this level. However, due to the previously mentioned divergences, most of the current commercial or open-source EDA tools developed for CMOS logic cannot be directly applied to the SFQ logic, especially for routing optimization in layout synthesis. This paper presents a hybrid Josephson transmission line (JTL) and passive transmission line (PTL) routing framework for SFQ logic, which fixes hold violations and optimizes critical path delay using a specialized global and detailed routing algorithm. The proposed routing tool comprises four parts: the data preparation, global routing, detailed routing and hybrid route widgets generation. More precisely, the routing environment for parallel computation is prepared by clustering and sorting the routing queue in the data preparation step. In the global routing step, the router calculates the best locations for each path by different types of routing layers. In the detailed routing step, the design timing is optimized based on the results of the global routing phase. After all the paths are optimized, the hybrid route widgets generator analyzes the grid nodes and produces JTL and PTL widgets. By applying this routing tool, seven testbench circuits are successfully routed with no hold violations, and with an operating frequency of up to 74.18 GHz, based on the proposed SIMIT-Nb03 [7, 8] technology and its advanced version referred to as

This work was supported by the Strategic Priority Research Program of Chinese Academy of Sciences (Grant No. XDA18000000), Shanghai Science and Technology Committee (Grant No. 21DZ1101000), the National Natural Science Foundation of China (Grant No. 62171437 and 92164101). *(Corresponding author: Jie Ren).*

Shucheng Yang, Xiaoping Gao, Ruoting Yang are with the Shanghai Institute of Microsystem and Information Technology, Chinese Academy of Science, Shanghai 200050, China and also with the CAS Center for Excellence in Superconducting Electronics (CENSE), Shanghai 200050, China.

Jie Ren, Zhen Wang are with the Shanghai Institute of Microsystem and Information Technology (SIMIT), Chinese Academy of Science (CAS), Shanghai 200050, China, with the CAS Center for Excellence in Superconducting Electronics (CENSE), Shanghai 200050, China and also with the Univerisity of Chinese Academy of Science Beijing, Beijing 100049, China (e-mail: jieren@mail.sim.ac.cn).

SIMIT-Nb04 in this paper, which is under development. The process consumes less than 688 s for each circuit. In addition, the framework is able to be compatible with advanced processes in the future, as long as the process design kit (PDK) data are imported.

The remainder of this paper is organized as follows. Section II introduces the previous works, standard cell library, timing constraints and placement and clock tree structure. Section III presents the proposed routing framework and algorithms. Section IV details the experimental results and their analysis. Finally, the conclusions are drawn in section V.

## II. BACKGROUND

### A. Previous work

Several SFQ routing algorithms exist [9-14]. In most of the cases, the logic gates in their standard cell library are designed have the same height. Moreover, due to the pipeline structure of the SFQ circuits, applying a row-based placement is usually a better choice. Most of the existing routing algorithms use a fixed routing structure (i.e. a fixed location of JTLs and PTLs) for connecting the clock tree and logic cells [9,10,14], or use the JTL-only [11] and PTL-only [12] routing strategies. Furthermore, the use of PTL for multi-fanout and short-distance interconnections is very limited, which is quite different from metal wires in CMOS logic. Using a hybrid routing methodology may improve the overall circuit performance, in order to save the layout space and optimize the maximum clock frequency.

### B. Standard Cell Library

#### 1) Logic Gates

Standard cells based on SIMIT-Nb03 [7, 8] and SIMIT-Nb04 do not typically have the same height, as mentioned in the previous section for JTL wiring and manufacturing reasons. All the logic gates are designed based on a unit pitch size. For instance, the sizes of the AND gate, OR gate and NOT gate are 2 units x 2 units, while that of the D flip-flop is 2 units x 1 unit. Note that the unit pitch size can be changed in different process versions. All the ports of logic gates are distributed around the block edge at the middle of the pitch (cf. Fig.1.).

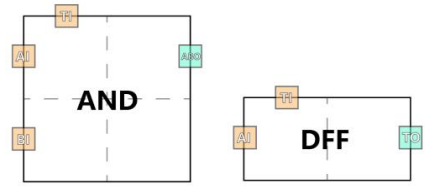

**Fig. 1.** Examples of logic gates layout in the SIMIT-Nb03 and SIMIT-Nb04 standard cell library.

In order to reduce the intrinsic delay of a logic gate, the ports are not connected with a PTL driver or receiver. If the PTL is applied in the Routing phase, the drivers and receivers should be placed in advance.

#### 2) Route units

The SIMIT-Nb03 and SIMIT-Nb04 standard cell library has two route units: Josephson Transmission Line (JTL) and Passive Transmission Line (PTL).

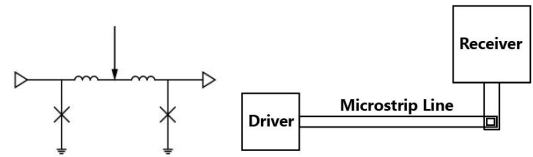

**Fig.2.** Schematics of (a) JTL (b) PTL

JTL is similar to a buffer. It provides a more stable transmission of SFQ pulses. However, it has a larger delay than PTL. The simplest JTL shown in Fig.2 (a) comprises two Josephson Junctions. When an SFQ pulse arrives at the input port, the first Junction is triggered. It then generates a new SFQ pulse to switch the second Junction, and the transmission process continues, and therefore the signal is delivered. A part of the JTL layouts in the SIMIT-Nb03 and SIMIT-Nb04 standard cell library is shown in Fig.3. The first two rows are referred to as long-JTL. They consist of 2 junctions. However, their layout is longer than normal. The last 4 layouts in the last row are standard size JTL containing 3 or 4 junctions. These JTL layouts allow a wide range of delay unit choices for fixing hold violations in the routing procedure.

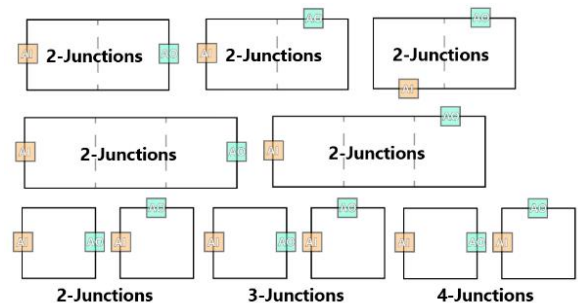

**Fig.3.** Examples of JTL layout in the SIMIT-Nb03 and SIMIT-Nb04 standard cell library.

PTL is similar to the metal wire. However, it requires a driver and receiver to connect with other SFQ cells at the beginning and end. A microstrip line is the core part of PTL, which can transmit an SFQ pulse at the speed of almost one-third of the light velocity. In this study, the unit time delay of PTL is approximately one-tenth of a standard 2-JJ JTL. However, similar to a JTL, the driver and receiver have a relatively larger delay. Only when the wire length satisfies equation (1), the PTL can have a shorter overall delay than JTL of the same length. In the considered standard cell library, this minimum length is 5 units.

$$t_{drv} + t_{rec} + (l - l_{drv} - l_{rec}) * t_{msl} \leq l * t_{jtl} \quad (1)$$

$$l \geq \frac{t_{drv} + t_{rec} - (l_{drv} + l_{rec}) * t_{msl}}{t_{jtl} - t_{msl}} \quad (2)$$

The design of the logic gates and route units will remain the same in the future advanced process. The only difference between the processes is the number of metal layers for power, clock and signal wiring.

*C. Timing constraints*

Similar to other digital logics, the timing constraints of the SFQ logic can be illustrated with a simple model, shown in Fig.4. Most of the SFQ logic gates are clocked. Therefore, a pair of connected SFQ logic gates follows the same timing constraints as the D flip-flop pair (cf. Fig.4). The timing constraints for the SFQ circuits can then be classified into three categories.

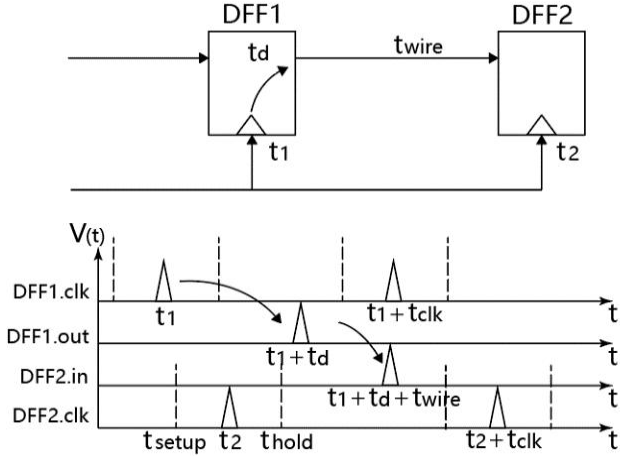

**Fig.4.** Timing constraints model.

1) **Hold time constraints**

The hold time constraints denote the length of a time window that allows an input signal to arrive after the clock arrives. These constraints are expressed as:

$$t_1 + t_{delay} + t_{wire} \geq t_2 + t_{hold} \quad (3)$$

where $t_{delay}$ and $t_{hold}$ are respectively intrinsic values of DFF1 and DFF2 that are not changeable in the optimization, $t_1$ and $t_2$ are respectively clock arrival times for DFF1 and DFF2, and $t_{wire}$ is the wire delay between DFF1 and DFF2.

The three variables can be optimized in order to fix the hold time violations in the routing process, which is a priority in the timing optimization phase of routing in this paper. The design will not correctly perform if a path fails due to hold time violations. Note that there are no other external means to fix these kinds of violations.

2) **Setup constraints**

The setup constraints are the time window which allows an input signal to arrive before the clock arrives. This constraint defines the maximum clock frequency value. It is expressed as:

$$t_1 + t_{delay} + t_{wire} \leq t_2 + t_{clk} - t_{setup} \quad (4)$$

For a specific path, the maximum of $t_{clk}$ can be determined by the other five variables. As mentioned in the hold time constraints, $t_1$, $t_2$ and $t_{wire}$ are optimizable. They are used to fix hold violations in priority. By combining the setup time constraints and hold time constraints, an optimization problem can be obtained:

$$\min t_{clk} \quad (5)$$
$$\text{s.t. } t_1 + t_{delay} + t_{wire} \geq t_2 + t_{hold} \quad (6)$$
$$t_1 + t_{delay} + t_{wire} \leq t_2 + t_{clk} - t_{setup} \quad (7)$$

By solving this optimization problem, the best $t_{clk}$ can be obtained and considered as the optimization target.

3) **I/O constraints**

The I/O timing constraints for a single SFQ circuit are independent of the previous constraints. The only requirement for a single circuit is that all the inputs should arrive either before the clock or after the clock, while it is better to be closer. This can be expressed as:

$$\min |t_{clk} - t_{input}| \quad (8)$$
$$\text{s.t. } t_{clk} > t_{input} \quad (9)$$

or

$$\min |t_{clk} - t_{input}| \quad (10)$$
$$\text{s.t. } t_{clk} < t_{input} \quad (11)$$

Ideally, the constraints (10) ~ (11) are preferred since they follow the same rules as the timing constraints inside the circuits. However, this consumes a large number of layout resources to delay the input signal. In addition, this is not an efficient solution in other designs that may use this circuit as a macro (cf. Fig.5). A practical method consists in following the first constraints and minimizing the difference between $t_{clk}$ and $t_{input}$. It is much easier to fix the timing violations when connecting two circuits, since only a slight delay should be added to a critical signal path.

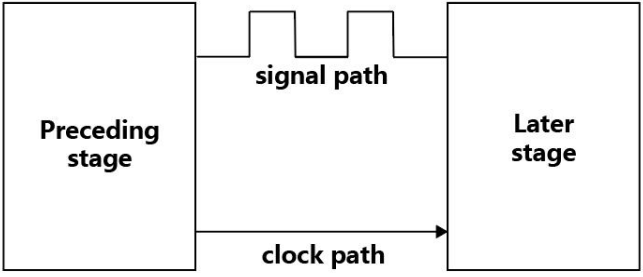

**Fig.5.** Schematic of interconnections between circuits

*D. Placement and Clock Tree Structure*

As previously mentioned, the clocked SFQ logic gates lead to the gate-level pipelined circuit structure. Bit-array is one of the best placement strategies. However, due to space limitations, the placement algorithm is not discussed in this paper. For a clock tree schema, the most recent works focus on the H-tree and its enhanced version HL-tree for SFQ logic [15]. In the experiments, the HL-tree and concurrent flow structure are used to minimize the size of the gigantic clock tree (cf. Fig.6.).

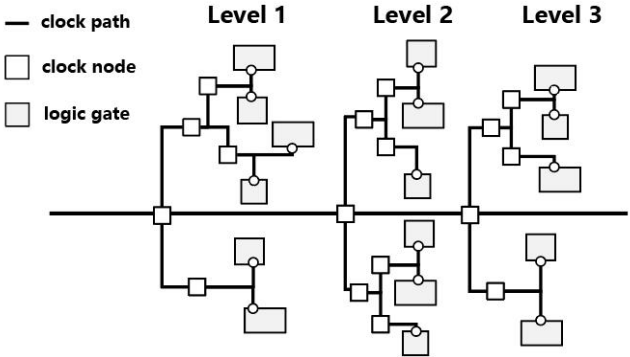

**Fig.6.** Example of placement and HL-clock tree structure.

### III. PROPOSED ALGORITHM

The proposed routing algorithm can be divided into four phases: the data preparation, global routing, detailed routing and hybrid route widgets generation. The proposed flow diagram is shown in Fig.8. The proposed routing strategy aims at improving the overall performance of the algorithm, while exploring the possibility of increasing the clock frequency of the SFQ design in a routable area.

In the data preparation phase, all the route prerequisites, such as the layout of logic gates, routable layer and synthesis result of a sizeable fan-out signal (in most cases, the clock and reset signal), should be obtained. The route map data is then processed together with a procedure of clustering and sorting route nets. The clustering procedure can improve the efficiency during the rip-up and reroute process in the global routing phase, and provide grouped data for multi-processing route calculation. However, for each routing group, the sequence of a routing queue determines the final results. Therefore, a better sorting strategy should be applied to each routing queue in order to improve the outcomes.

In the global routing phase, the primary task consists in connecting all the nets in the given route layer. In contrast to the CMOS routing algorithm, two attributes for the routing layer are defined: the JTL enabling layer and PTL enabling layer. For the JTL enabling layer, the routing search program follows the JTL path rules, including the routing direction constraints and generating JTL cross and splitters. The PTL enabling layer follows the PTL path rules with PTL routing constraints. These two types of routing layers are controlled by given processes. Currently, only when the routing search program cannot completely find a path on the bottom layer (which has to be the JTL enabling layer), the top layer (which may be the PTL enabling layer) will be activated.

The JTL and PTL enabling layers are defined as the attributes of a metal layer in the framework. For the global router, the JTL enabling layer rules have a priority over the PTL enabling layer rules (i.e. on the metal layer which is both a JTL enabling layer and PTL enabling layer). The global router will try JTL routing first, since a JTL path can be more easily optimized than a PTL path in the detailed routing phase. In the SIMIT-Nb03 process, two metal layers only exist for logic cells and wiring (cf. Fig.7.). M0 and M1 are defined as both JTL enabling layer and PTL enabling layer, in order to potentially use PTL for delay optimization in the detailed routing phase. In the SIMIT-Nb04 process, we have M2 and M3 for Micro-Strip Line (MSL) only. Therefore, M2 and M3 are defined as PTL enabling layer, while M0 and M1 remain the same. In this case, only when M2 and M3 are out of space, a PTL delay optimization on M0 and M1 is allowed in the detailed routing phase.

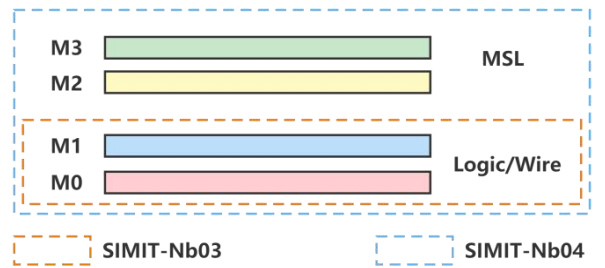

**Fig.7.** Metal layers for logic and wiring in SIMIT-Nb03 and SIMIT-Nb04.

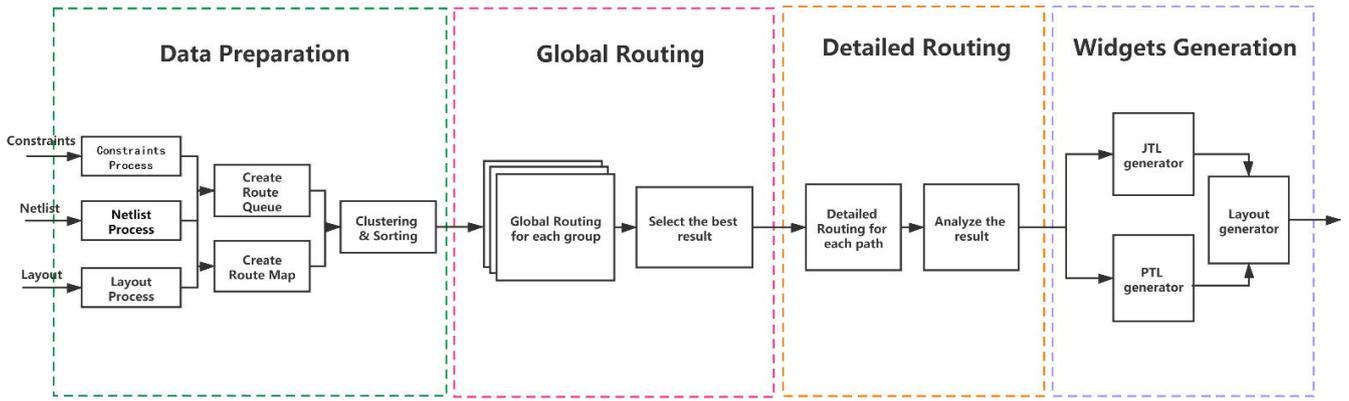

**Fig.8.** Flowchart of the routing process

Once there is no path for a net in the group, the rip-up and reroute procedure is performed, until the program reaches its exit condition.

The detailed routing phase aims at fixing up the hold violation and optimizing the path delay for all the paths. The hybrid JTL and PTL optimizer are applied with a random optimization strategy, in order to explore the optimization space in the proposed algorithms. In the beginning, the program adjusts all the clock tree nets in order to minimize the clock tree skew in each logic level. It then calculates all the clock arrival time, output port emits time, hold time window and setup time window. Given all the timing data, the hold time constraints and setup time constraints can be derived.

The optimizer can fix all the hold violations and shorten the critical path delay. After the first optimization cycle, the detailed router will try to perform an incremental optimization on the clock tree bridges and branches in order to improve the overall performance. In contrast to other existing routers, different delay units (i.e. 3 or 4-junction JTL) are first used to enlarge the path delay, rather than detouring the violated path, which possibly saves the routable area for other nets.

Finally, in the hybrid route widgets generation phase, the JTL layout and PTL layout widgets are generated. The generator analyzes the path information from detailed routing, including its coordinates and widgets type, then outputs a script file according to the user input option. By running the output script file in corresponding layout edit tools, the routed layout is transferred, and the following tasks, such as DRC and LVS, can be performed.

In the sequel, details about these three phases are provided.

*A. Data Preparation*

Before running the algorithm, a routing environment should be prepared. As previously mentioned, the input files consist of layout, netlist and large fan-out information. The layout information contains logic gates with gate model, origin coordinates and orientation. The program will then use these data to create a route map. The logic gates are added as the blocked area on corresponding layers for global routing calculations in the route map. Route coordinate pairs for global routing can also be obtained by combining the layout and interconnection data derived from the netlist. The large fan-out information is processed and converted into route coordinate pairs. It is then appended to the routing queue. Apart from large fan-out signals, the coordinate pairs are clustered for multi-processing calculation. The coordinates are the criterion for the clustering program. If one coordinate pair intersects with an existing group, it will be added to this group. Otherwise, the program will generate a new group. For each group, a margin enlarges the border. Typically, a margin of 5 units is enough for the proposed clustering program. The diagram of the clustering result is presented in Fig.9.

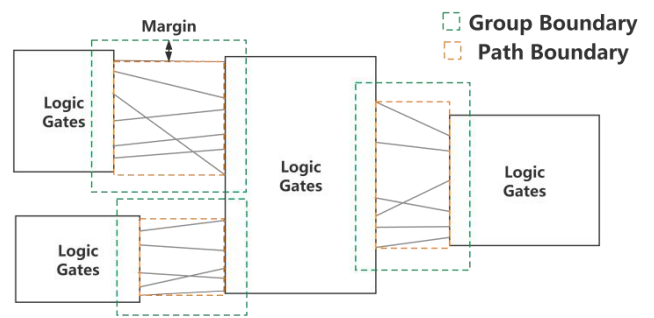

**Fig.9.** Diagram of the clustering result.

After clustering, the coordinate pairs in each group are sorted by fan-out of a net descending. The coordinate pairs having the same fan-out are sorted using the sum of Manhattan distance descending. However, the large fan-out signals are always the priority in the routing queue.

## B. Global Routing

In this phase, the used global router is based on a modified version of the A* algorithm [13]. The basic A* algorithm calculates the sum of the move cost and estimate cost. However, in the proposed global router, two new items of cost are used: corner cost and layer cost. The updated formulation is expressed as:

$$f_{cost} = g_{cost} + h_{cost} + c_{cost} + l_{cost} \qquad (12)$$

After finding the destination, the program selects a path of shortest length with the minimum number of corners and best location on the routing layer. All these conditions are essential for later processes. In addition, a searching bound box of 4 units margin (cf. Fig.10) is assigned before the searching program starts, which can shorten the time for searching a path that is failed. The algorithms used in this part are provided in Algorithm 1.1 and Algorithm 1.2.

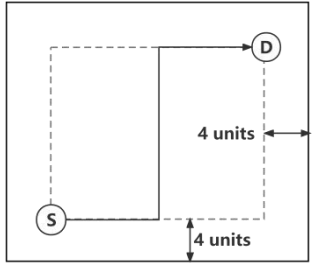

Fig.10. Route boundary for a single path.

**Algorithm 1.1:** Modified A* routing

1: **Input**: source, destination, Map
2: **Output**: path
3: /* Start path searching */
4: currLocation ← source
5: /* Create two list for path searching. openList is for nodes to be checked, and closedList is for nodes that are inoperable*/
6: openList,closedList ← initialize()
7: **Append** currLocation **to** openList
8: **while** True **do**
9:   // Find the fast position (minimum $f_{cost}$) in the openList, as current location
10:   currLocation ← getFastPosition(openList)
11:   **if** currLocation is **None** ||
12:     currLocation is the destination **then**
13:     // Exit whenthe destination is found or no invalid
14:     path
15:     **break**
16:   **Append** currLocation **to** closedList
17:   openList.pop(currLocation)
18:   // Find the adjacent positions of the current location
19:   posList ← getAdjacentPositions() // Algorithm 1.2
20:   // Calculate the cost of each node
21:   **for** pos **in** posList **do**
22:     **if** pos **in** closedList **then**
23:       **continue**
24:     **else**
25:       H,G,C,L ← calculateCost(pos,map,destination)
26:     **if** pos **in** openList && pos.G > G **then**
27:       pos.G ← G
28:       pos.F ← G+H+C+L
29:       pos.preLoc ← currLocation
30:     **else**
31:       **Append** pos **to** openList
32:       pos.G ← G
33:       pos.F ← G+H+C+L
34:       pos.preLoc ← currLocation
35:   **end**
36: **end**
37: /* Record the path start, from destination to source*/
38: **while** currLocation is not **None do**
39:   **Append** currLocation **to** path
40:   currLocation ← currLocation.preLoc
41: **end**
42: **return** path

Algorithm 1.2 presents a function for finding adjacent positions in Algorithm 1.1. The constraints mentioned above and below, are performed in this function.

**Algorithm 1.2:** Obtain the Adjacent Position

1: **Input**: currLocation, searchBound, pathDict, Map
2: **Output**: Positions
3: /* Get moving offsets under constraints*/
4: **if** currLocation on JTL layer **then**
5:   // Corner case. Only one direction is allowed.
6:   **if** currLocation is a corner on other path **then**
7:     offsets ← Corner direction
8:   // Cross case. Only straight direction is allowed.
9:   **else if** Map[currLocation]==1 **then**
10:     offsets ← Straight directions
11:   **else if** PTL channel opened **then**
12:     offsets ← 6-way directions
13:   **else if** currLocation.G>4 **then**
14:     offsets ← 4-way directions, vertical first
15:   **else**
16:     offsets ← 4-way directions, horizontal first
17: **else if** currLocation on PTL layer **then**
18:   **if** layer.index is odd **then**

```
19:        offsets←Verticle directions
20:     else
21:        offsets←Horizontal directions
22:  /* Get adjacent positions*/
23:  adjPositions←initialize()
24:  for offset in offsets do
25:     nextLocation←currLocation+offset
26:     if nextLocation is out of searchBound then
27:        continue
28:     else if map[nextLocation]>=2 then
29:        continue
30:     else
31:        Append nextLocation to adjPositions
32:  end
33:  return adjPositions
```

Besides these constraints, the routed paths and blockages on the route map also affect the global router. For the JTL enable layer, 4-way moving directions are applied. Only when enabled, in cases such as no valid path for this coordinate pair and PTL layer exists, the additional choice of moving upwards or downwards through different layers are allowed. The capacity of a JTL enable layer is set to 2 for the SIMIT-Nb03 process technology. The map is encoded as follows:

0: nothing on the node.
1: one path on the node, which means another path can go through this node using JTL-cross.
2: two paths are crossed on this node, or a splitter is placed on it.
3: Blockage.

If the net is a single fan-out, the path coordinates with corresponding encoded values are added to the route map after each path is routed. For multi-fanout cases, this process is performed after all the paths of this net are routed and refined (cf. Fig.11).

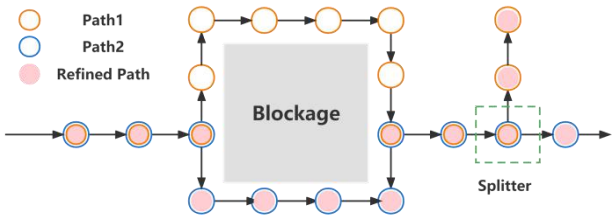

**Fig.11.** Diagram of the multi-fanout path refinement in the global routing phase.

For the PTL enable layer, the constraints are much easier. Similar to the CMOS metal wire, the PTL layer is divided into horizontal and vertical layers. On a specific PTL layer, only vertical or horizontal directions are allowed in terms of the layer index. In addition, a PTL path should start from and end at a JTL layer, as the driver and receiver should be placed on a JTL layer. However, the global router will not route a PTL unless the JTL layer is used up, since in most of the cases, the PTL path can hardly achieve multi-fanout. Performing a path replacement for delay optimization in the detailed routing phase, is a better way for using PTL.

The overall global routing control flow is shown in Fig.12. If a failed path exists in a route group, the rip-up and reroute program will be triggered. All the map data of routed paths in this group will be cleaned up, and the failed paths will be added to the front of the route queue. A restart routing is then performed for all the paths. The rip-up and reroute program will also determine whether the route queue after resorting is invalid, i.e. the resorted route queue is similar to a previous one in the history. In this case, the program will exit and report the failed path to the designers. Finally, the global router will select a routed path result with the least failed paths and highest layout area utilization for each group.

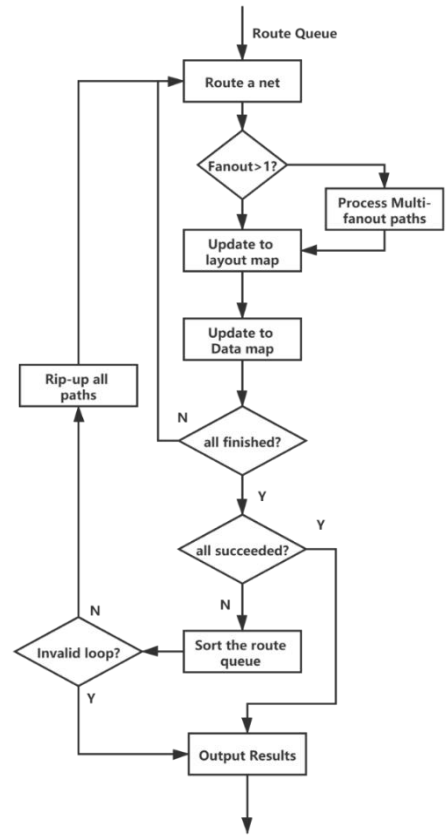

**Fig.12.** Flowchart of the global routing algorithm.

*C. Detailed Routing*

The proposed detailed router aims at fixing and optimizing the timing for the routed paths from the global routing. In order to optimize the critical paths, different types of JTLs and PTLs in the cell library are used to achieve flexible and accurate timing adjustment, rather than detouring routed paths for fixing timing violations or using the rip-up and reroute strategy. For both clock tree and signal path optimization, a multi-fanout path optimizer which can fix hold time violations and sequentially shorten critical paths, is developed.

For a multi-fanout net, the following scheme is used to explain the algorithm. After global routing, the multi-fanout path is naturally divided by the splitters. From the destinations to the source, the path segments are sorted by their index size ascending. In the example, the optimize sequence is PathSeg1, PathSeg2, PathSeg3, PathSeg4, PathSeg2.3, PathSeg1.2.3 and PathSeg1.2.3.4. This optimization strategy can ensure the minimum influence to other paths. For a single fanout net, since only one path exists in the group, this strategy is also applied.

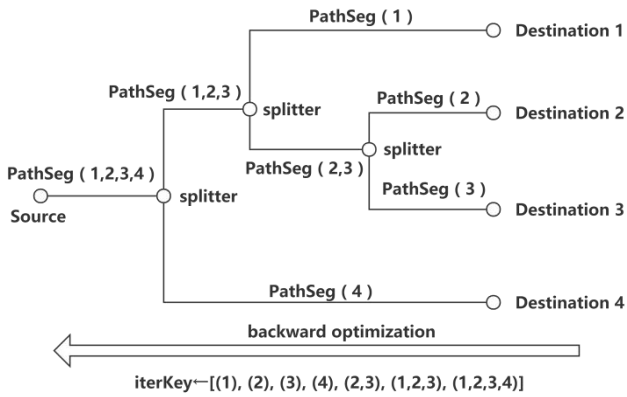

**Fig.13.** Scheme of a multi-fanout path.

The multi-fanout path optimizer first analyzes the encoded value of the path segments group on all the route maps, in order to derive the changeable widgets. The latter are all widgets except those containing splitters, JTL cross, driver, receiver and MSL. Before optimization, the target timing and current timing of each routed path are calculated and then sent to the optimizer. For each widget, the estimate timing uses the standard delay of a 2-junction JTL for the JTL path, and standard delay of a unit MSL cell for the PTL path.

For each path segment in the sorted list, the priority is to fix hold violation using the 3 or 4-junction JTLs in order to randomly replace the changeable widgets, then updating the timing. Only when all the paths are out of hold violation in the path segment, the optimizer starts shortening the path segment by replacing the changeable widgets with long-distance JTLs. In addition, if PTL is allowed, the optimizer will first try to use PTL replacement. Note that in the shorten operation, the maximum number of path widgets that can be shortened is calculated based on the worst timing constraints in this path segment according to Eq.(3). This means that no more hold violations should be produced in this operation.

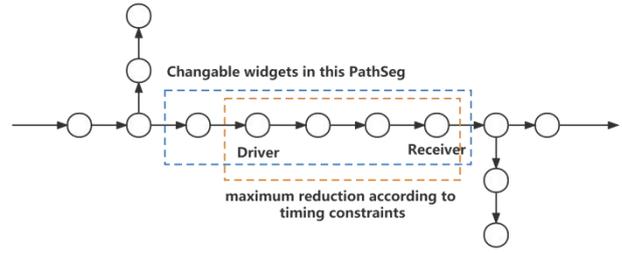

**Fig.14.** Diagram of path segments optimization. The circles represent the path coordinates (widgets) in the layout map. PTL driver and receiver are placed in terms of layout map condition and maximum reduction of this PathSeg.

**Algorithm 2:** Multi-Fanout Path Optimizer

1: **Input**: targets, paths, map, options
2: **Output**: residueList, newPath
3: /* Store the path coordinates information (location, path index and number of junctions) into a hash table */
4: posDict←getPosInfo(paths,map)
5: /* For each path, calculate the current residue between the path length and target length */
6: residueList←calculateResidue(paths,targets)
7: /* Create the pathseg sequence (cf. Fig.13) */
8: iterKey←getPathSegIndex(posDict)
9: /* Fix hold violations */
10: **for** i **in** iterKey **do**
11:   **if** all residue **in** residueList >=0 **then**
12:     // all the paths are out of hold violation, exit
13:     **break**
14:   // Select all pos in posDict that pos.index==i
15:   posList←getPosList(posDict,i)
16:   **for** pos **in** posList **do**
17:     **if** pos.junctions<4 **then**
18:       pos.junctions+=1
19:     // Update the residueList
20:     updateResidue(residueList,pos,variance)
21:     **if** all residue **in** residueList >=0 **then**
22:       // all the paths are out of hold violation, exit
23:       **break**
24:   **end**
25: **end**
26: /* optimize delay */
27: **for** i **in** iterKey **do**
28:   **if** all residue **in** residueList <=1 **then**

```
29:         // all the paths have already been optimized,
30:         exit
31:         break
32:     // find the pathSeg in posDict and map (cf. Fig.14)
33:     pathSeg←getPathSeq(posDict,i)
34:     if PTL layer is enabled then
35:         // Firstly, try ptl routing for optimization
36:         // Get the maximum reduction in this group
37:         reduction←getReduction(residueList,pathSeg)
38:         // find a proper location for driver and receiver
39:         source,dest←getPTLRouteInfo(reduction, map)
40:         // reroute the path in target PTL layer
41:         PTLpath←AStarRoute(source,dest,ptlMap)
42:         // replace the original path with PTLpath
43:         updatePathInfo(pathSeg, PTLpath)
44:         // update the residueList
45:         update(residueList, pathSeg, variance)
46:     // Secondly, try long-jtl for optimization.
47:     reduction←getReduction(residueList, pathSeg)
48:     // find the locations for long-JTL
49:     longJTLPath←shortenPath(reduction, map)
50:     // update the optimization result
51:     updatePathInfo(pathSeg, PTLpath)
52:     update(residueList, pathSeg, variance)
53: end
```

Based on the previously presented multi-fanout path optimizer, the detailed router performs the optimization in four phases:

(I) For concurrent-flow HL trees, the optimization range of each sub-tree is calculated by the optimizer. However, no optimization results are presently applied. According to the optimization range, the target timing of each sub-tree is calculated by level descending. In other words, if the clock arrival time of the final level is n, the target timing of the previous sub-tree should ideally be no less than n. After clock tree optimization, the timing skew of each sub-tree is almost balanced, depending on the layout condition. Simultaneously, the clock arrival time of each node is marked.

(II) The output emit time, hold time window and setup time window are computed by giving the timing information from the clock tree optimization and the cell library. The target timing and current timing of all the routed signal paths are then calculated. With all the prepared requirements, the multi-fanout path optimizer can work on the routed signal path, in order to fix the hold time violation and shorten the critical path.

(III) Retiming the clock tree. After signal path optimization, the timing conditions are updated. For each logic level, the minimum value of $t_{wire} - t_{hold}$ is calculated. In other words, the clock bridge between levels can be optimized under this constraint. If the constraint of one logic level is below zero, the optimizer will shorten the clock tree bridge between the previous and current logic level, in order to eliminate the hold violations. Otherwise, the optimizer will lengthen this bridge to help reduce the maximum delay. Afterwards, for each instance, the optimizer will adjust the clock tree branch connected to it, based on the timing of the current instance and the instance connected to it in the next logic level (cf. Fig.15 (d)~(e)). This is due to the fact that moving the clock arrival time of the current instance affects the output emit time and the data arrival time of the next instances. The maximum value for delaying the clock arrival time of an instance is expressed as:

$$min\ (t_{curr\_slack} - t_{next\_slack})/2 \qquad (13)$$
$$s.t.\ t_{curr\_slack} = t_{curr\_wire} - t_{curr\_hold} \qquad (14)$$
$$t_{next\_slack} = t_{next\_wire} - t_{next\_hold} \qquad (15)$$

The program will gather all the related timing information, find the best target for all the clock tree branches in one clock sub-tree, then call the multi-fanout path optimizer again in order to optimize this clock sub-tree, while ensuring that no more hold violations are created in this phase.

(IV) If IO routing is applied, the input stage is optimized based on the constraints in Section II.D (3).

Once the detailed routing is completed, the timing conditions of all the logic gates and the layout size are re-calculated for performance analysis reports. In addition, the path widgets type is marked with its coordinate for the layout generation.

*D. Hybrid Route Widgets Generation*

Based on the optimization results obtained from the detailed router, the hybrid route widgets generator starts by analyzing the path coordinates with its corresponding widget type. For the JTL widget, a long-JTL is created in terms of its start and end coordinates, while the standard size JTL, splitter and cross are created based on the input directions, output directions and number of junctions. For the PTL paths, the first and last coordinate together with the route directions, are used to create the driver and receiver, respectively. The middle part of a PTL is used to create micro-strip line widgets and vias in terms of its widget coordinate, layer, input and output directions.

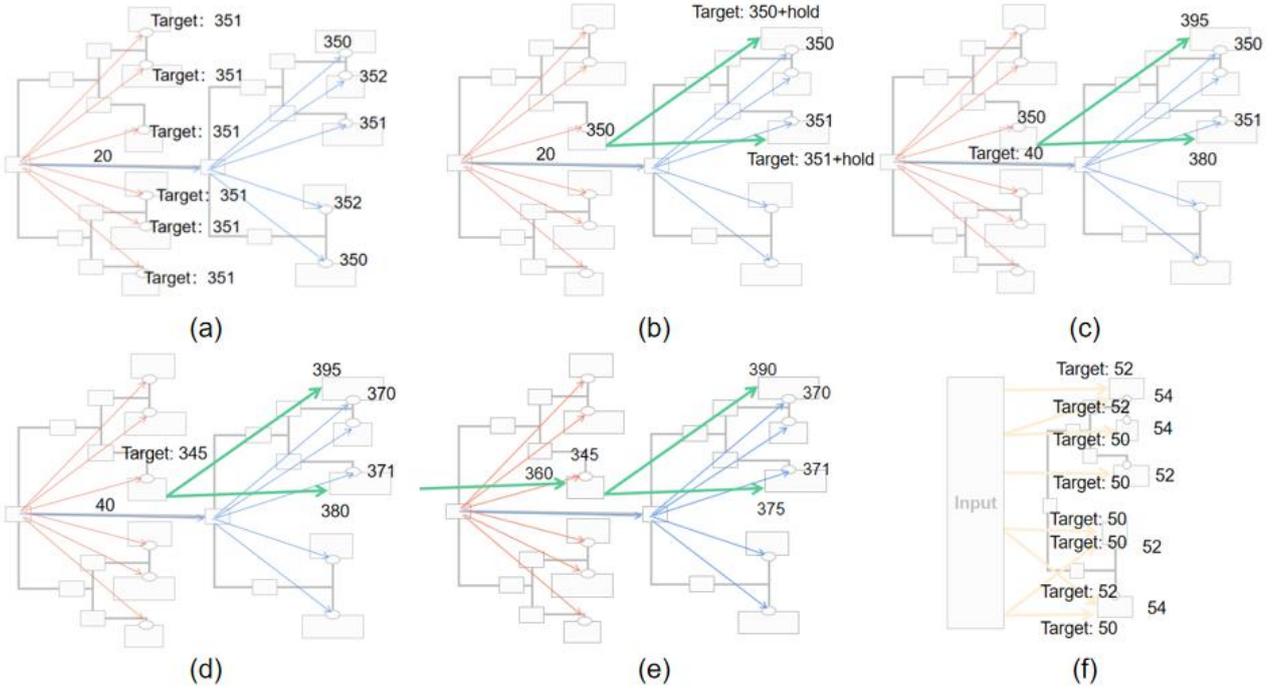

**Fig.15.** (a) Phase I: optimizing the clock tree. (b) Phase II: optimizing the signal path. (c) Phase III: retiming the clock bridge. (d) Phase III: retiming the clock sub-tree. (e) Final result for one net. (f) Phase IV: optimizing the input stage.

All the widget information will be compiled and outputted to the script file for corresponding layout edit tools. The designers can then run the script and generate the routed layout for future use. Due to the fact that a path is optimized based on its timing constraints in detailed routing, it is possible that the widget generator presents an all-PTL path, a hybrid path or an all-JTL path after compilation. Examples of the possible generation results are presented in Fig.16.

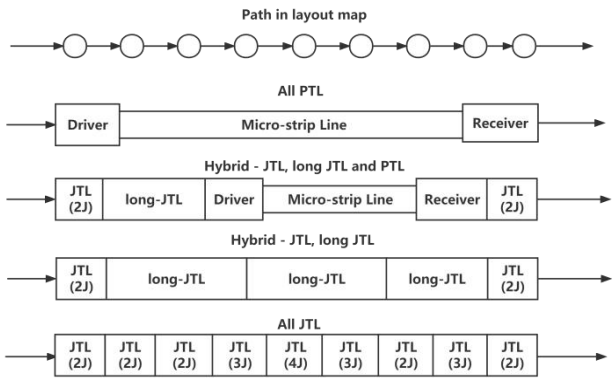

**Fig.16.** Possible generation results for a path segment after compilation (not exhausted).

### IV. EXPERIMENTAL RESULTS AND ANALYSIS

The hybrid routing algorithm is implemented using Python and C++ on the server, with Intel Xeon CPU E5-2698 @ 2.3Ghz and 252.2 GB Memory. Some circuits from ISCAS c-series and previous SFQ designs performed by our team, are selected as test cases, using the SIMIT-Nb03 and SIMIT-Nb04 technologies.

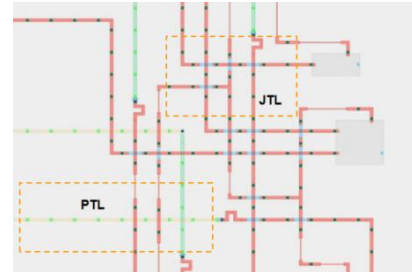

**Fig.17.** JTL and PTL hybrid symbol layout.

For the SIMIT-Nb03 technology, the framework is used to verify the functionality and fix the hold violations, since only two metal layers exist, so that the delay optimization (especially PTL usage) is limited. The SIMIT-Nb04 technology is mainly used in order to show the ability to optimize the post-routing clock frequency. In practice, the layout placement still needs to be manually adjusted, due to the fact that the available placement and clock tree synthesis tools are not efficient enough. The best experimental results of each testbench circuits based on SIMIT-Nb04, are summarized in Table I.

Theoretically, if the placement and clock tree synthesis result is ideal, the routing optimization result should be close to the limitation of the given standard cell library. For the SIMIT-Nb03 and SIMIT-Nb04 standard cell library, the timing limitation is derived from:

$$T_{max\_clk} = max(T_{hold} + T_{setup}), \qquad (16)$$

which is 13.2 ps or 75.6 GHz, including a safe value for hold and setup constraint. In general, the maximum expected clock frequency is almost 56.2 GHz, since it is limited by the timing constraints of the worst gate. Note that in our library, the worst one is the NOT gate.

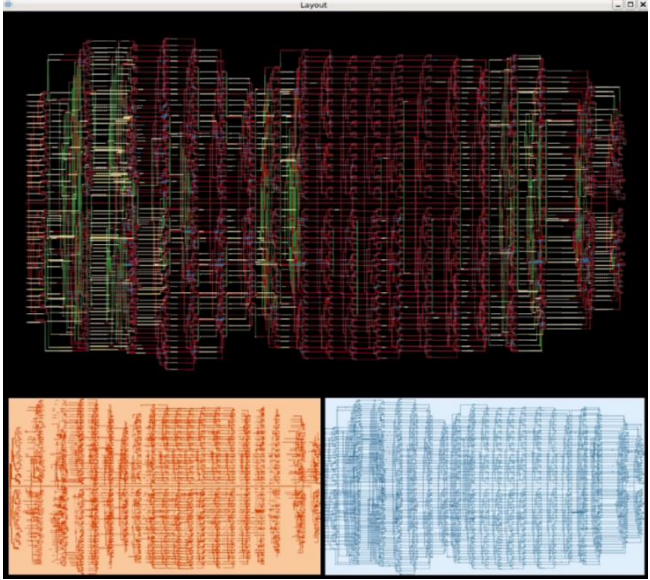

**Fig.18.** C499 (a) Post-routing layout. (b) Junction density. (c) Route density.

It can be seen from Table I that C17 and SR-array are simple enough to achieve the ultimate clock frequency after detailed routing. For C499 and other complex circuits, the maximum clock frequency is restricted by the critical path, even though the optimizer will prioritize it. However, the maximum clock frequency may increase if a better detailed placer is applied, due to the lack of a well-optimized layout placement. In terms of fixing hold violations, the proposed framework is able to eliminate all the violations in the given testbench circuits.

In addition, the waveform of logic synthesis netlist and post-layout netlist are compared for post-layout verification. Fig.19 shows that after the routing optimization, the functionality of the testbench circuit is matched with its logic synthesis result.

Besides the less-optimized placement, the layout size and wire length may be restricted due to the used stricter PTL design rules based on the manufacturing and testing results, such as wider MSL, larger size of via, corner and interface, for example.

## V. CONCLUSION

This paper presents a hybrid JTL and PTL routing framework for large-scale SFQ logic design automation. By combining the utilization of JTL and PTL, the proposed tool can accurately perform on the SIMIT-Nb03 and SIMIT-Nb04. It highly increases the maximum clock frequency and fixes almost all the hold violations. The global router of the proposed tool uses a modified A* algorithm for multi-layer and multi-type global routing, finds the best path coordinates and shape on both JTL layer and PTL layer. It then achieves the highest routing completion rate. The detailed router of the proposed tool uses a specially designed path optimizer and four-phases optimization framework for accurate timing adjustments.

Table I
Routing Results of ISCAS c-series, 4bit-KSA, 8bit-ALU, AES-SBOX and 64x64-SR array

| circuit | Pre-routing /post-routing junctions | Nets | Area (mm$^2$) | Wire length (μm) | JTL/PTL usage (unit) | Pre-routing/ post-routing hold violations | Worst slack (ps) | Pre-routing/ post-routing frequency (GHz) | Run time (s) |
|---|---|---|---|---|---|---|---|---|---|
| C17 | 234/683 | 54 | 0.323 | 9.2E3 | 161/54 | 5/0 | 9.40 | 26.71/72.54 | 0.052 |
| C499 | 29274/90365 | 3400 | 63.79 | 1.51E6 | 4.1E4/2.2E4 | 178/0 | 5.24 | 1.28/18.83 | 142.8 |
| C1908 | 51884/184563 | 6073 | 126.2 | 2.16E6 | 5.2E4/2.3E4 | 1356/0 | 3.81 | 1.03/12.35 | 687.9 |
| 4bit-KSA | 2394/6374 | 314 | 5.367 | 1.12E5 | 2.1E3/1.6E3 | 7/0 | 8.46 | 6.24/38.12 | 2.563 |
| 8bit-Alu | 7938/19438 | 767 | 16.34 | 1.47E5 | 4.6E3/9.3E2 | 86/0 | 6.40 | 4.78/24.72 | 6.243 |
| AES-SBOX | 10254/44167 | 1926 | 38.46 | 8.46E5 | 1.8E4/1.1E4 | 114/0 | 9.66 | 3.98/28.74 | 16.38 |
| 64x64-SR array | 33748/291203 | 12415 | 205.46 | 8.32E6 | 1.5E5/1.1E5 | 7552/0 | 10.24 | 55.2/74.35 | 63.52 |

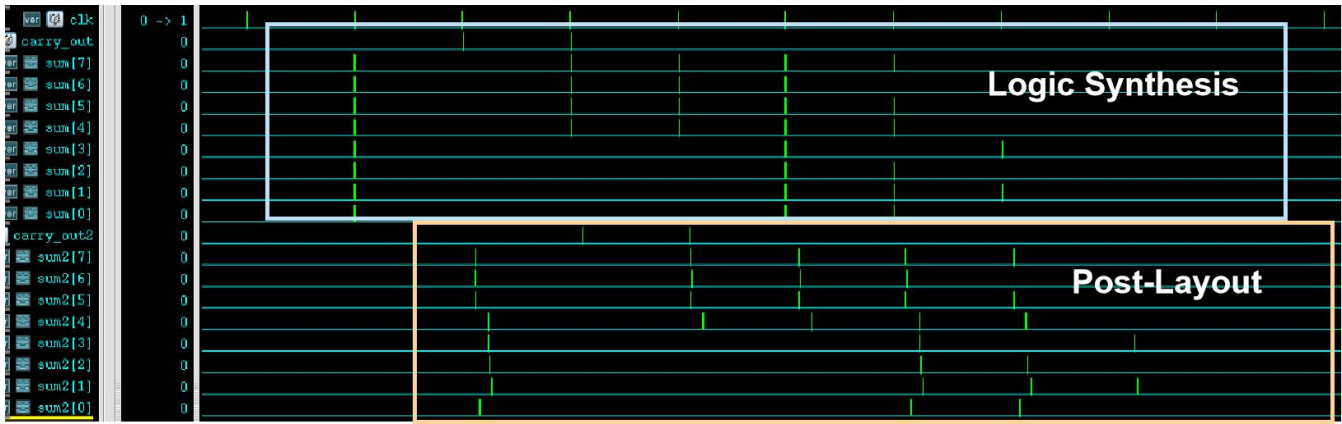

**Fig.19.** Waveform comparison between logic synthesis netlist and post-layout netlist (8bit-ALU).

Finally, the experimental results demonstrate that the proposed tool successfully optimizes several testbench circuits at clock frequencies up to 74.18 GHz, with no hold violations and an execution time less than 688 s.

ACKNOWLEDGMENT

The authors would like to thank Ling Xin for inspiring routing algorithms as reference, and Xi Gao, Qi Qiao for providing PTL design rules and testing data.